\documentclass[epjc,aps,showpacs,preprintnumbers,amsmath,nofootinbib,amssymb]{revtex4}
\usepackage[dvips]{epsfig}

\newcommand{\nc}[1]{\newcommand{#1}}

\nc{\its}[1]{\itshape #1 \upshape}
\nc{\mc}[3]{\multicolumn{#1}{#2}{#3}}

\nc{\bc}{\begin{center}}
\nc{\ec}{\end{center}}
\nc{\ig}[1]{\bc \includegraphics{#1} \ec}

\nc{\bo}[1]{\mbox{\boldmath \( #1 \! \! \)  \unboldmath}}
\nc{\be}{\begin{eqnarray}}
\nc{\ee}{\end{eqnarray}}
\nc{\bew}{\begin{eqnarray*}}
\nc{\eew}{\end{eqnarray*}}
\nc{\bs}{\begin{subeqnarray}}   
\nc{\es}{\end{subeqnarray}}     
\nc{\nnn}{\nonumber \\}
\nc{\f}[2]{\frac{#1}{#2}}
\nc{\td}[2]{\f{d #1}{d #2}}
\nc{\pd}[2]{\f{\partial #1}{\partial #2}}
\nc{\suli}{\sum\limits}
\nc{\proli}{\prod\limits}
\nc{\ili}{\int\limits}
\nc{\sr}[2]{\stackrel{#1}{#2}}
\nc{\dps}{\displaystyle}
\nc{\ket}[1]{\left| #1 \right>}
\nc{\bra}[1]{\left< #1 \right|}
\nc{\bracket}[2]{\left< #1 \right| \left. \! #2 \right>}
\nc{\norm}[1]{\left\| #1 \right\|}
\nc{\lndm}[1]{\pd{^{#1} \ln{\det{M}}}{\mu^{#1}}}
\nc{\pdmm}[1]{M^{-1} \pd{^{#1} M}{\mu^{#1}}}
\nc{\pdm}{M^{-1}\pd{M}{\mu}}
\nc{\trac}[1]{\mbox{Tr}\left(#1\right)}

\def\lsim{\raise0.3ex\hbox{$<$\kern-0.75em\raise-1.1ex\hbox{$\sim$}}}
\def\gsim{\raise0.3ex\hbox{$>$\kern-0.75em\raise-1.1ex\hbox{$\sim$}}}

\begin{document}
\mbox{} \hfill BI-TP 2005/52\\[0mm]
\mbox{} \hfill BNL-NT-05/49\\[0mm]
\mbox{} \hfill TKYNT-05-30
\title{The isentropic equation of state of 2-flavor QCD}

\author{S. Ejiri$^{\rm a}$, F. Karsch$^{\rm b,c}$, E. Laermann$^{\rm c}$ and 
C. Schmidt$^{\rm b}$}

\address{
$^{\rm a}$Department of Physics, The University of Tokyo,
\small Tokyo 113-0033, Japan\\
$^{\rm b}$Physics Department, Brookhaven National Laboratory, 
Upton, NY 11973, USA \\
$^{\rm c}$Fakult\"at f\"ur Physik, Universit\"at Bielefeld, D-33615 Bielefeld,
Germany
}

\date{\today}

\begin{abstract}
Using Taylor expansions of the pressure obtained 
previously in studies of 2-flavor QCD at non-zero 
chemical potential we calculate expansion coefficients for the energy and 
entropy densities up to ${\cal O}(\mu_q^6)$ in the quark chemical potential. 
We use these series in $\mu_q/T$ to determine lines of constant entropy 
per baryon number ($S/N_B$) that characterize the expansion of dense 
matter created in heavy ion collisions.
In the high temperature regime these lines are found to be well
approximated by lines of constant $\mu_q/T$. 
In the low temperature phase, however, the quark chemical 
potential is found to increase with decreasing temperature. This is 
in accordance with resonance gas model calculations. Along the lines
of constant $S/N_B$ we calculate the energy density and pressure. 
Within the accuracy of our present analysis we find that
the ratio $p/\epsilon$ for $T>T_0$ as well as the softest point of the 
equation of state, $(p/\epsilon)_{min}\simeq 0.075$,  show no significant
dependence on $S/N_B$. 

\end{abstract}


\maketitle

\section{Introduction}
\label{intro}

Recently studies of QCD thermodynamics on the lattice have successfully 
been extended to non-vanishing quark (or baryon) chemical potential using 
Taylor expansions \cite{eos,eos6} as well as reweighting techniques \cite{Fodor} 
around the limit of vanishing quark chemical potential ($\mu_q=0$) 
and analytic continuations of numerical
calculations performed with an imaginary chemical potential \cite{lombardo}. 
The complementary approach at fixed baryon number \cite{redlich,karsch}
has also been used recently in simulations of 4-flavor QCD with light 
quarks \cite{deForcrand}.
These calculations yield the additional contribution to the pressure
that arises from the presence of a net excess of baryons over anti-baryons
in a strongly interacting medium. We will further explore here 
the approach based on a Taylor expansion of the pressure and make use of our 
earlier calculations for 2-flavor QCD \cite{eos6} to determine also the energy 
and entropy densities at non-vanishing baryon chemical potential. 

In a heavy ion collision a dense medium is created which after thermalization
is expected to expand without further generation of entropy ($S$) and with
fixed baryon number ($N_B$) or, equivalently, with fixed quark number $N_q=3N_B$. 
During the isentropic expansion the ratio
$S/N_B$ thus remains constant\footnote{Here $N_B$ denotes the
net baryon number, i.e. the excess of baryons over anti-baryons.}.
The cooling of the expanding system then is
controlled by the equation of state on lines of constant $S/N_B$.
From a knowledge of the energy density and pressure at non-vanishing 
quark chemical potential we can calculate trajectories in the $\mu_q$-$T$ 
phase diagram of QCD that correspond to constant $S/N_B$ and can determine 
the {\it isentropic equation of state} on these trajectories.

This paper is organized as follows. In the next section we summarize the
basic setup for calculating Taylor expansions for bulk thermodynamic
observables and present a calculation of the Taylor expansion coefficients 
for energy and entropy densities at non-vanishing quark chemical potential  
up to ${\cal O}(\mu_q^6)$. We use these results in
Section III to determine lines of constant $S/N_B$ and study the 
temperature dependence of pressure and energy density along these lines.
Our conclusions are given in section IV. 

\section{Taylor expansion of pressure, energy and entropy density}
\label{taylor}

The analysis we are going to present in this paper is based on 
numerical results previously obtained in simulations of 2-flavor QCD on
lattices of size $16^3\times 4$ with an improved staggered fermion action
\cite{eos,eos6}. In these calculations Taylor expansion coefficients
for the pressure have been obtained up to ${\cal O}(\mu^6)$ for a fixed 
bare quark mass value ($\hat{m}=0.1$) which at temperatures close to
the transition temperature ($T_0$) corresponds to a still quite large
pion mass of about 770~MeV. However, in particular at temperatures above
the QCD transition temperature the remaining quark mass dependence of 
thermodynamic observables is nonetheless small as  deviations from the
massless limit are controlled by the quark mass in units of the temperature,
which is a small number. 

We closely follow here the approach and notation used in Ref.~\cite{eos6}. We 
start with a Taylor expansion for the pressure in 2-flavor QCD for 
non-vanishing quark chemical potential $\mu_q$ (and vanishing isospin
chemical potential, $\mu_I\equiv 0$),
\begin{equation}
\frac{p}{T^4}\equiv \frac{1}{VT^3}\ln Z(T,\mu_q)=
\sum_{n=0}^\infty c_n(T,m_q) \left(\frac{\mu_q}{T}\right)^n\quad ,
\label{eq:p}
\end{equation}
with 
\begin{equation}
c_n(T,m_q) =\frac{1}{n!}\frac{1}{VT^3}{{\partial^n \ln Z(T,\mu_q)}
\over{{\partial(\mu_q/T)^n}}}\biggr\vert_{\mu_q=0}
\quad .
\label{eq:cn}
\end{equation}
Here we explicitely indicated that the Taylor expansion coefficients
$c_n$ depend on temperature as well as the quark mass\footnote{We
assume that the thermodynamic limit has been taken so that we
can ignore any volume dependence of thermodynamic quantities.
For an aspect ratio $TV^{1/3}=4$ and the quark mass value used in
our calculation finite volume effects are expected to be small.}.
In the context of lattice calculations it is more customary
to think of this quark mass dependence in terms of a dependence on
an appropriately chosen ratio of hadron masses characterizing {\it 
lines of constant physics}. We thus may replace $m_q$ by a ratio
of pseudo-scalar (pion) and vector (rho) meson masses,
$m_q\equiv (m_{PS}/m_V)^2$.

Due to the invariance of the partition function under exchange of
particle and anti-particle the Taylor expansion is a series in 
even powers of $\mu_q/T$; expansion coefficients $c_n$ vanish for
odd values of $n$.
From the pressure we immediately obtain the quark number density,
\begin{equation}
\frac{n_q}{T^3} = \frac{1}{VT^3}\frac{\partial\ln Z(T,\mu_q)}{\partial
\mu_q/T}\biggl|_{T} = 
\sum_{n=2}^\infty n c_n(T,m_q) \left({\mu_q\over T}\right)^{n-1}\quad .
\label{eq:nq}
\end{equation}
Using standard thermodynamic relations we also can calculate the difference
between energy density ($\epsilon$) and three times the pressure,
\begin{equation}
\frac{\epsilon - 3p}{T^4} = \sum_{n=0}^\infty c'_n(T,m_q) 
\left({\mu_q\over T}\right)^n\quad ,
\label{eq:e3p}
\end{equation}
where 
\begin{equation}
c'_n(T,m_q) =T \frac{{\rm d} c_n(T,m_q)}{{\rm d} T}  \quad .
\label{eq:cne3p}
\end{equation}
Combining Eqs.~\ref{eq:p},~\ref{eq:nq}, and \ref{eq:e3p} we then obtain Taylor 
expansions for the energy and entropy densities,
\begin{eqnarray}
\frac{\epsilon}{T^4} &=& \sum_{n=0}^\infty \left(3 c_n(T,m_q) +
c'_n(T,m_q)\right) \left({\mu_q\over T}\right)^n\quad ,\nonumber \\
\frac{s}{T^3} \equiv \frac{\epsilon +p-\mu_q n_q}{T^4}
&=& \sum_{n=0}^\infty \left( (4-n) c_n(T,m_q) +
c'_n(T,m_q)\right) \left({\mu_q\over T}\right)^n\quad .
\label{eq:es}
\end{eqnarray}
The expansion coefficients $c_n(T,m_q)$ have been calculated in 
2-flavor QCD at several values of the temperature and for a fixed
value of the bare quark mass, $\hat{m}=0.1$ \cite{eos6}. 
We note that the bare coupling $\hat{m}$ introduces an implicit
temperature dependence if $\hat{m}$ is kept fixed while 
the lattice cut-off is varied. The latter controls the temperature of the system on
lattices of temporal extent $N_\tau$, {\it i.e.} $T^{-1}=N_\tau a$. The difference
between derivatives taken at fixed $\hat{m}$ and fixed $m_q$,
however, is negligible at high temperature, where
the quark mass dependence of bulk thermodynamic quantities is 
${\cal O}((m_q/T)^2)$ and, of course, vanishes at all temperatures
in the chiral limit as this also defines a line of constant physics. 
We thus approximate the derivative, Eq.~\ref{eq:cne3p},
by a derivative taken at fixed $\hat{m}$ which becomes exact in the chiral
limit. These derivatives are evaluated using finite difference approximants.
Alternatively we may express derivatives of Taylor expansion coefficients
with respect to $T$ in terms
of derivatives with respect to the bare lattice couplings $\beta\equiv 6/g^2$ and
$\hat{m}$ and 
rewrite Eq.~\ref{eq:cne3p} as,
\begin{eqnarray}
c'_n(T,m_q) =-a \frac{{\rm d}\beta}{{\rm d} a} \frac{\partial
c_n(T,m_q)}{\partial \beta} -
a\frac{{\rm d}\hat{m}}{{\rm d} a} \frac{\partial 
c_n(T,m_q)}{\partial \hat{m}} 
  \quad .
\label{eq:cne3pb}
\end{eqnarray}
Here the two lattice $\beta$-functions, $a{\rm d}\beta/{\rm d} a$
and $a{\rm d}\hat{m}/{\rm d} a$, have to be determined by demanding
that the temperature derivative is taken along lines of constant 
physics \cite{milc,AliKhan}. The $\beta$-function controlling the
variation of the bare quark mass with the lattice cut-off, 
$a{\rm d}\hat{m}/{\rm d} a$, is proportional
to the bare quark mass $\hat{m}$ and thus vanishes in the chiral 
limit \cite{milc}.
For small quark masses the first term in Eq.~\ref{eq:cne3pb} thus gives 
the dominant contribution to $c'_n(T,m_q)$.
We have evaluated this first term in Eq.~\ref{eq:cne3pb} for $n=2$
and find that this approximation for $c'_2$
agrees within errors with the calculation of $c'_2$
from finite difference approximants. Only close to $T_0$ we find 
differences of the order of $10\%$.
 
\begin{figure}[tb]
\begin{center}
\begin{minipage}[c]{15.5cm}
\begin{center}
\epsfig{file=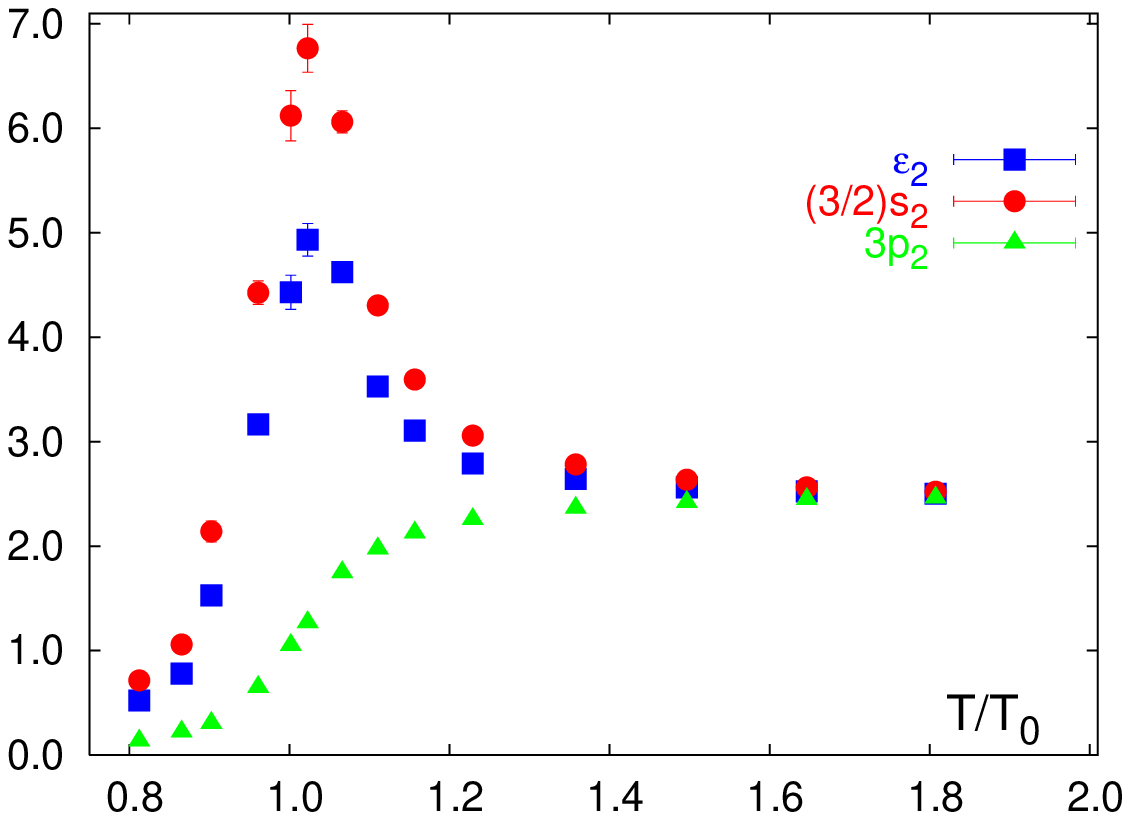, width=7.5cm}\hspace{0.2cm}
\epsfig{file=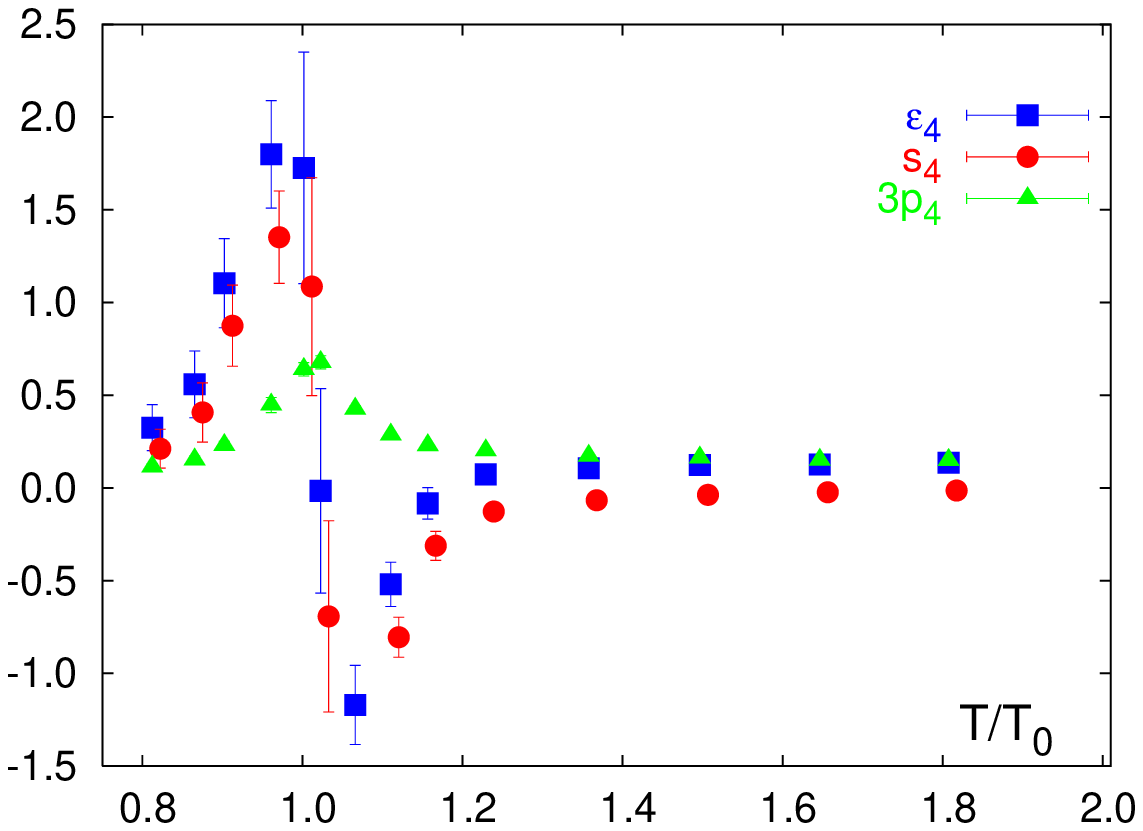, width=7.5cm}

\epsfig{file=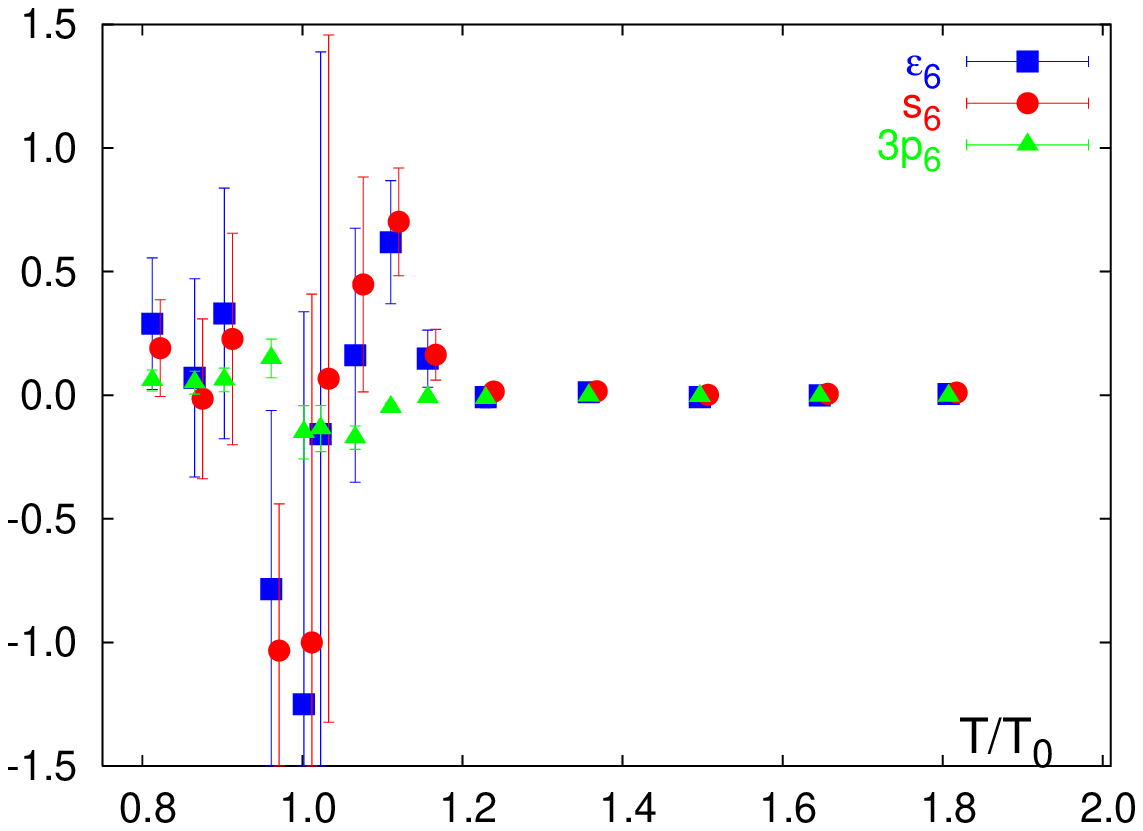, width=7.5cm}
\end{center}
\end{minipage}
\end{center}
\caption{The $2^{nd}$, $4^{th}$ and $6^{th}$ order Taylor expansion 
coefficients for pressure ($p_n\equiv c_n$), energy density ($\epsilon_n$) 
and entropy density ($s_n$) as functions of $T/T_{0}$.}
\label{fig:es}
\end{figure}

The expansion coefficients $c_n(T,m_q)$ have been calculated in \cite{eos6}
for $\hat{m}=0.1$ at a set of 16 temperature values in the interval 
$T/T_0\in [0.81,1.98]$.
From this set of expansion coefficients we have calculated the partial
derivatives
$c'_n$ at temperature $T$ by averaging over finite difference approximants 
for left and right derivatives at $T$. With this we have constructed the 
expansion coefficients for the energy and entropy densities,
\begin{eqnarray}
\epsilon_n &\equiv& 3 c_n(T,m_q) + c'_n(T,m_q) \; , \nonumber\\
s_n &\equiv& (4-n) c_n(T,m_q) + c'_n(T,m_q) \; .
\label{coef}
\end{eqnarray}
Results for the $2^{nd}$, $4^{th}$
and $6^{th}$ order expansion coefficients of energy and entropy
densities 
obtained in the approximation discussed above are shown in Fig.~\ref{fig:es}. 
Also shown
in this figure are the corresponding expansion coefficients for the pressure
($p_n\equiv c_n$).
For temperatures larger than $T/T_0\simeq 1.5$ all expansion
coefficients satisfy quite well the ideal gas relations,
$\epsilon_{n}^{SB}=3p_{n}^{SB}$ and $\epsilon_{2}^{SB} = 3 s_{2}^{SB}/2$. 
As noted already in Ref.~\cite{eos6} results
for the $2^{nd}$ and $4^{th}$ order expansion coefficients are  
close to those of an ideal Fermi gas which describes
the high temperature limit for the energy density of 2-flavor QCD,
\begin{equation}
\frac{\epsilon^{SB}_F}{T^4} = 
\frac{7\pi^2}{10}+
3 \left( \frac{\mu_q}{T}\right)^2 +
\frac{3}{2\pi^2} \left( \frac{\mu_q}{T}\right)^4 \; . 
\label{epsilon_SB}
\end{equation} 
Nonetheless, quantitative statements on the approach to ideal gas behavior may
be at present premature. Deviations from the continuum result clearly arise 
in our current analysis which has been performed on lattices with 
a lattice spacing in units of the inverse temperature given by
$aT\equiv 1/N_\tau=1/4$. Although cut-off effects are reduced due the 
usage of an improved action the approach to the continuum limit 
has to be analyzed further through calculations on lattices with larger 
temporal extent $N_\tau$.

\begin{figure}[tb]
\begin{center}
\begin{minipage}[c]{14.5cm}
\begin{center}
\epsfig{file=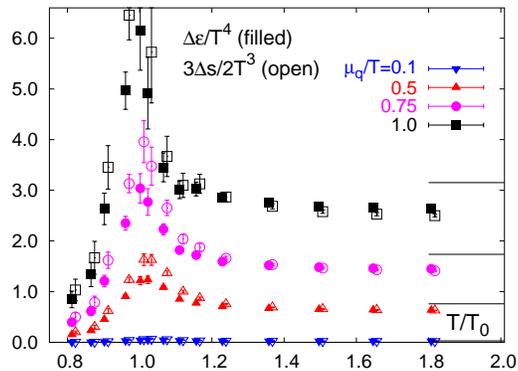, width=7.5cm}
\end{center}
\end{minipage}
\end{center}
\caption{The $\mu_q$ dependent contribution to  energy density (filled symbols) 
and entropy density (open symbols) calculated in $4^{th}$ order Taylor expansion. 
Data points for the entropy density  have been shifted slightly for better 
visibility. Horizontal lines show the corresponding ideal gas values for
the energy density.
}
\label{fig:thermo}
\end{figure}

Although errors on the $6^{th}$ order expansion coefficients are large
it is apparent that subsequent orders in the expansion decrease in 
magnitude thus making the expansions for energy and entropy density 
well behaved for $\mu_q/T\lsim 1$. In fact, the $6^{th}$ order contribution
to $\epsilon /T^4$ and $s/T^3$ never exceeds 10\% for $\mu_q/T < 0.9$.
We will see below that this is the regime of interest for a discussion of 
the thermodynamics of matter formed in heavy ion collisions at the SPS and
RHIC and even is appropriate for the energy range to be covered by the
AGS and the
future FAIR facility at GSI/Darmstadt. Results for the $\mu_q$-dependent 
contribution to the energy and entropy densities, $\Delta\epsilon/T^4 \equiv
\epsilon/T^4 -\epsilon_0$ and $\Delta s/T^3 = s/T^3-s_0$,  obtained in $4^{th}$ order
Taylor expansion are shown in 
Fig.~\ref{fig:thermo} for some values of $\mu_q/T$. 

\section{Lines of constant entropy per baryon number}
\label{entropy}

When discussing the equation of state at finite baryon number density
we have to eliminate the quark chemical potential in the thermodynamic
quantities introduced in the previous section. We may do this by
demanding that a thermodynamic quantity, for instance the quark
number density, stays constant. 
We will use here a prescription more appropriate for discussing the
thermodynamics of matter created in relativistic heavy ion collisions. 
After equilibration the dense medium created in such a  
collision will expand along lines of constant entropy per baryon. It then
is of interest to calculate thermodynamic quantities along such 
isentropic lines. 

We thus should first analyze how the quark chemical potential
changes along lines of constant $S/N_B$. It is readily seen that 
in an ideal quark-gluon gas these lines are defined by $\mu_q/T=$~const, 
\begin{equation}
\frac{S}{N_B} = 3\; {\frac{37\pi^2}{45}+ 
\left(\frac{\mu_q}{T}\right)^2
\over
\frac{\mu_q}{T}+
\frac{1}{\pi^2} \left(\frac{\mu_q}{T}\right)^3
}\quad .
\label{idealSNq}
\end{equation}
In the zero temperature limit, however, the resonance gas reduces
to a degenerate Fermi gas of nucleons and the chemical potential
approaches a finite value to obtain finite baryon number and entropy,
{\it i.e.} $\mu_q/T \sim 1/T$.  

\begin{figure}[tb]
\begin{center}
\begin{minipage}[c]{15.8cm}
\begin{center}
\epsfig{file=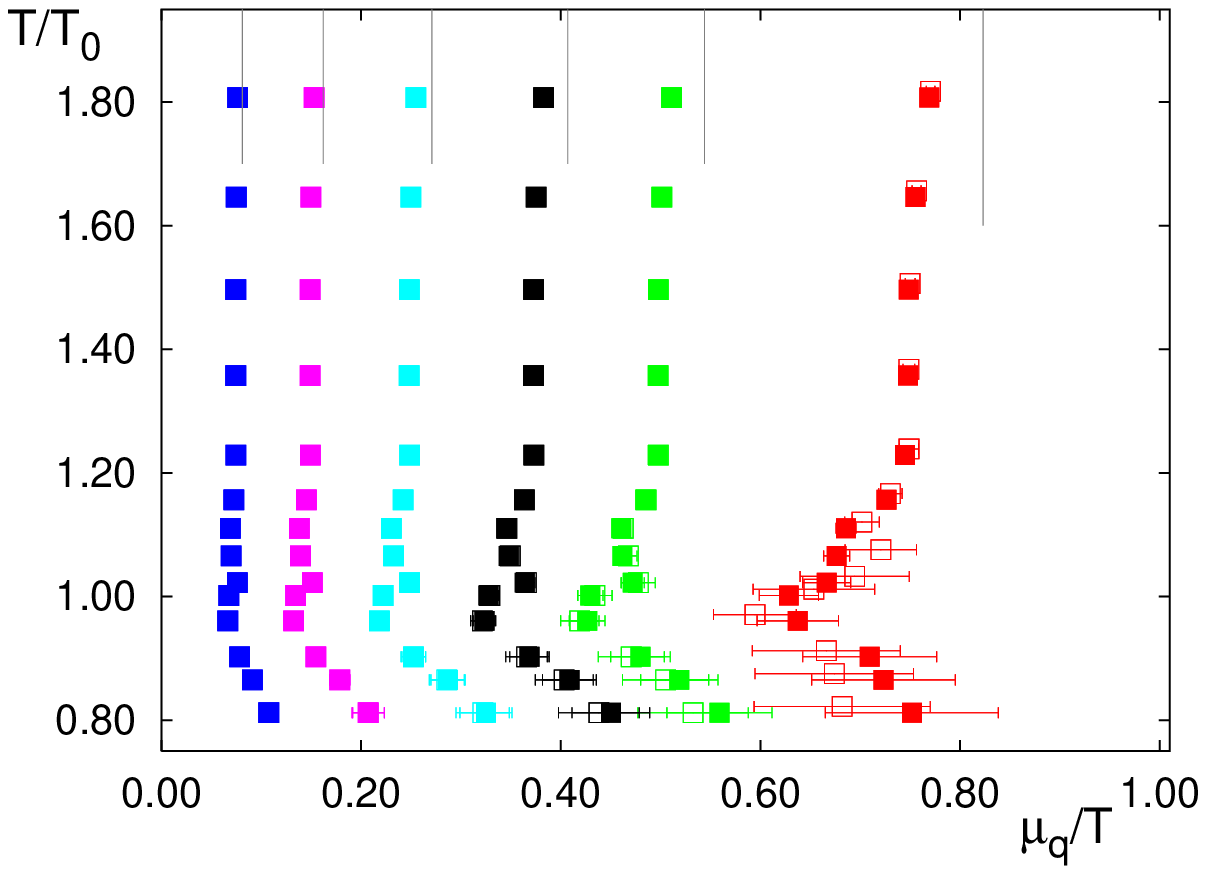, width=7.7cm}\hspace{0.2cm}
\epsfig{file=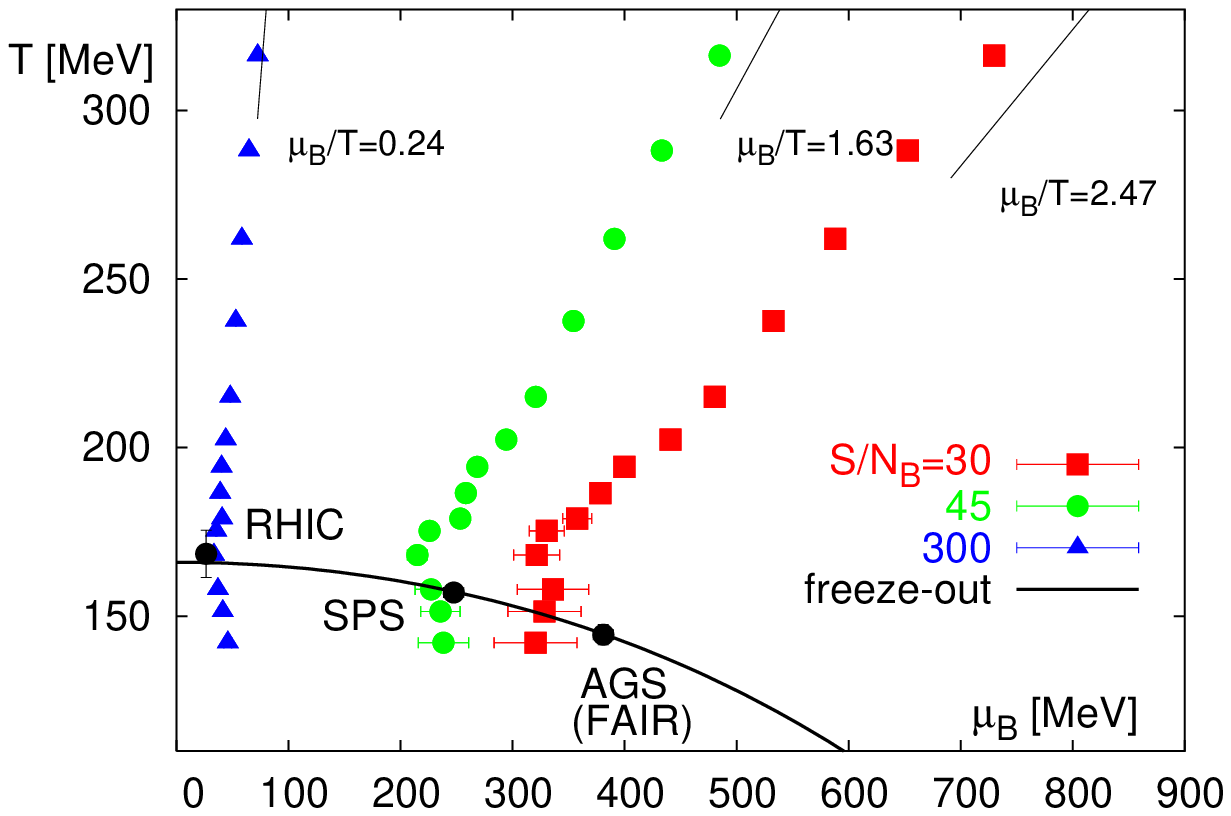, width=7.7cm}
\end{center}
\end{minipage}
\end{center}
\caption{Lines of constant entropy per quark number versus $\mu_q/T$ (left)
and in physical units using $T_0=175$~MeV to set the scales (right).
In the left hand figure we show results obtained using a $4^{th}$ (full symbols)
and $6^{th}$ (open symbols) order Taylor expansion of the pressure,
respectively. Data points correspond to $S/N_B = 300$, 150, 90, 60, 45, 30
(from left to right).
The vertical lines indicate the corresponding ideal gas 
results, $\mu_q/T = 0.08$, 0.16, 0.27, 0.41, 0.54 and 0.82 in decreasing 
order of values for $S/N_B$.
For a detailed description of the right hand figure see the discussion
given in the text.
}
\label{fig:iso}
\end{figure}

In Fig.~\ref{fig:iso} we show the values for $\mu_q$ needed to keep
$S/N_B$ fixed. These lines of constant $S/N_B$ in the $T$-$\mu_q$ plane
have been obtained by calculating the total entropy density for 2-flavor
QCD from Eq.~\ref{eq:es} using results for the pressure and energy density
calculated at $\mu_q=0$ \cite{peikert,peikertPhD} and the corresponding
$\mu_q$ dependent contributions shown in Fig.~\ref{fig:es}. The ratio
of $s/T^3$ and $n_q/T^3$ obtained in this way is then solved numerically
for $\mu_q/T$.

We find that isentropic expansion at high temperature indeed is well
represented by lines of constant $\mu_q/T$ down to temperatures close
to the transition, $T\simeq 1.2 T_0$. In the low temperature regime we 
observe a bending of the isentropic lines in accordance with the expected
asymptotic low temperature behavior. The isentropic
expansion lines for matter created at SPS correspond to $S/N_B  
\simeq 45$ while the isentropes at RHIC correspond to 
$S/N_B \simeq 300$. The energy range of the AGS which also corresponds
to an energy range relevant for future experiments at FAIR/Darmstadt
is well described  by $S/N_B \simeq 30$. These lines are shown in 
Fig.~\ref{fig:iso}(right) together with points characterizing the 
chemical freeze-out of hadrons measured at AGS, SPS and RHIC energies.
These points have been obtained by comparing experimental results
for yields of various hadron species with hadron abundances in a
resonance gas 
\cite{cleymans,redlich05}. The solid curve shows a phenomenological
parametrization of these {\it freeze-out data} \cite{redlich05}.
In general our findings for lines of constant $S/N_B$ are in good 
agreement with phenomenological model calculations that are
based on combinations of ideal gas and resonance gas equations of
state at high and low temperature, respectively \cite{shuryak,toneev}.

Our current analysis yields stable results for lines of constant $S/N_B$
also for temperatures $T\lsim T_0$
and values of the chemical potential $\mu_B \simeq 400$~MeV. This value of 
$\mu_B$ is larger than recent estimates
for the location of the chiral critical point in 2-flavor \cite{gavai_gupta} and 
(2+1)-flavor \cite{fodor_ccp} QCD. We thus may expect modifications to our
current analysis at temperatures below $T_0$ once we include higher orders
in the Taylor expansion and/or perform this analysis at smaller values of the
quark mass. 
 
\begin{figure}[tb]
\begin{center}
\begin{minipage}[c]{15.8cm}
\begin{center}
\epsfig{file=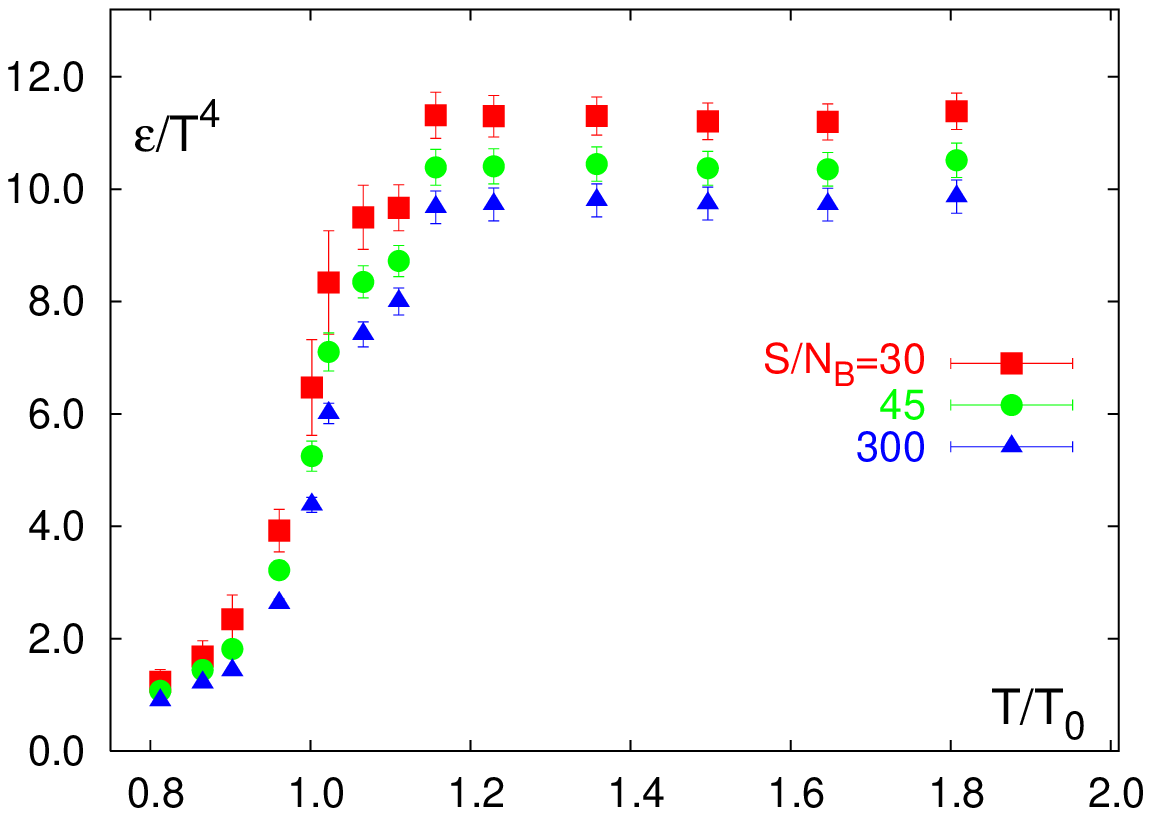, width=7.7cm}\hspace{0.2cm}
\epsfig{file=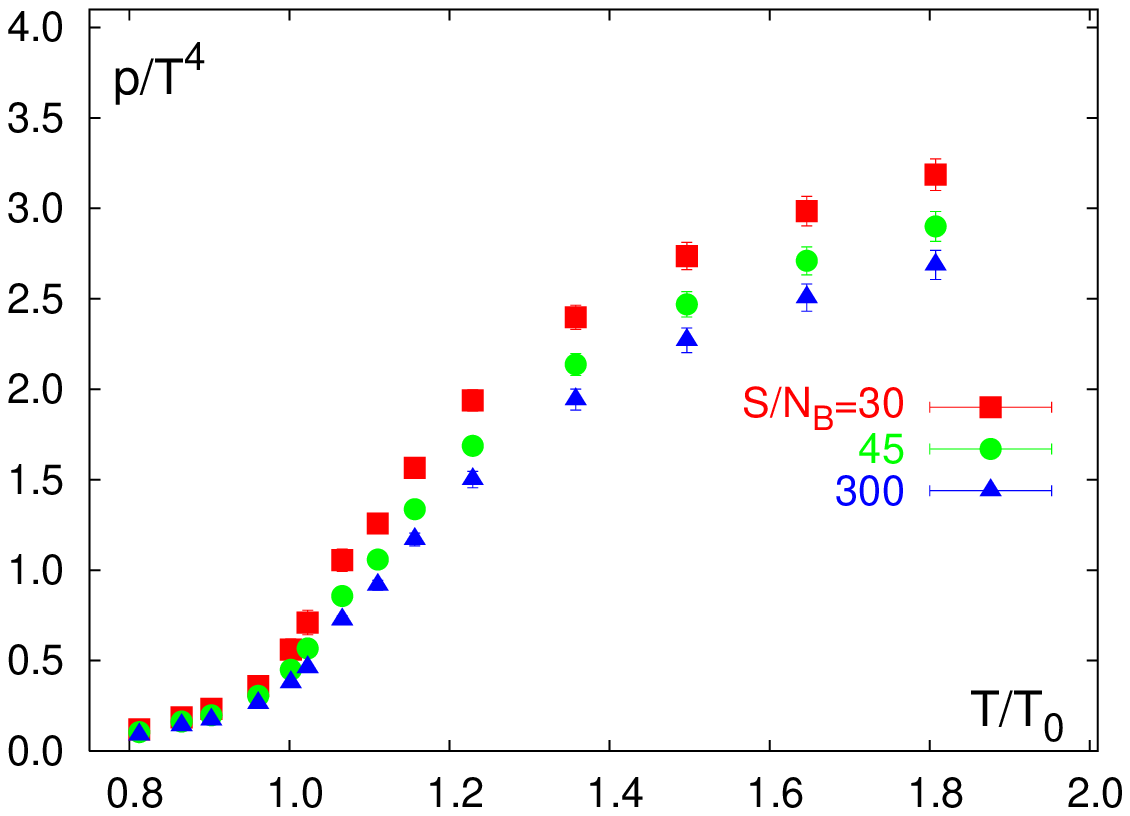, width=7.7cm}
\end{center}
\end{minipage}
\end{center}
\caption{The complete energy density (left) and pressure (right) 
evaluated on lines of constant $S/N_B$ using 
Taylor expansions up to $6^{th}$ order in the 
quark chemical potential. Shown are results for
$S/N_B=30,~45,~300$.
}
\label{fig:eos}
\end{figure} 

\begin{figure}[tb]
\begin{center}
\begin{minipage}[c]{15.8cm}
\begin{center}
\epsfig{file=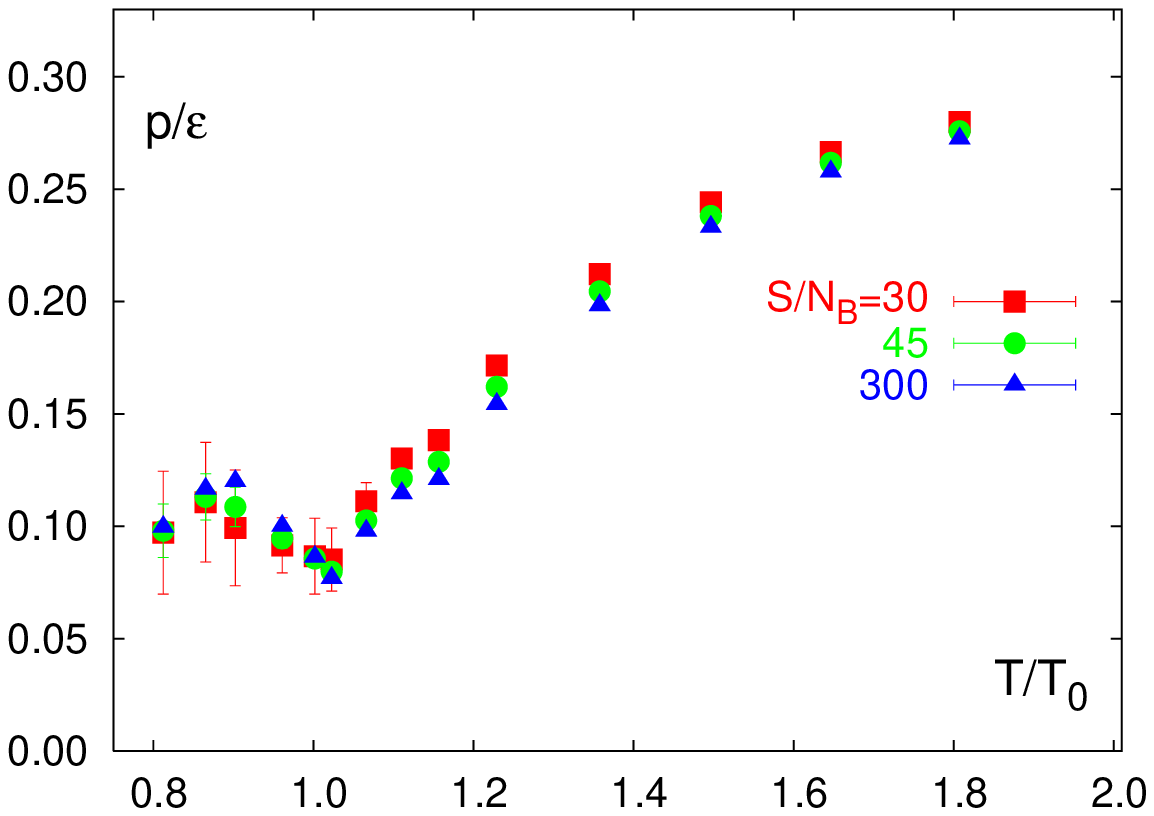,width=7.7cm}\hspace{0.2cm}
\epsfig{file=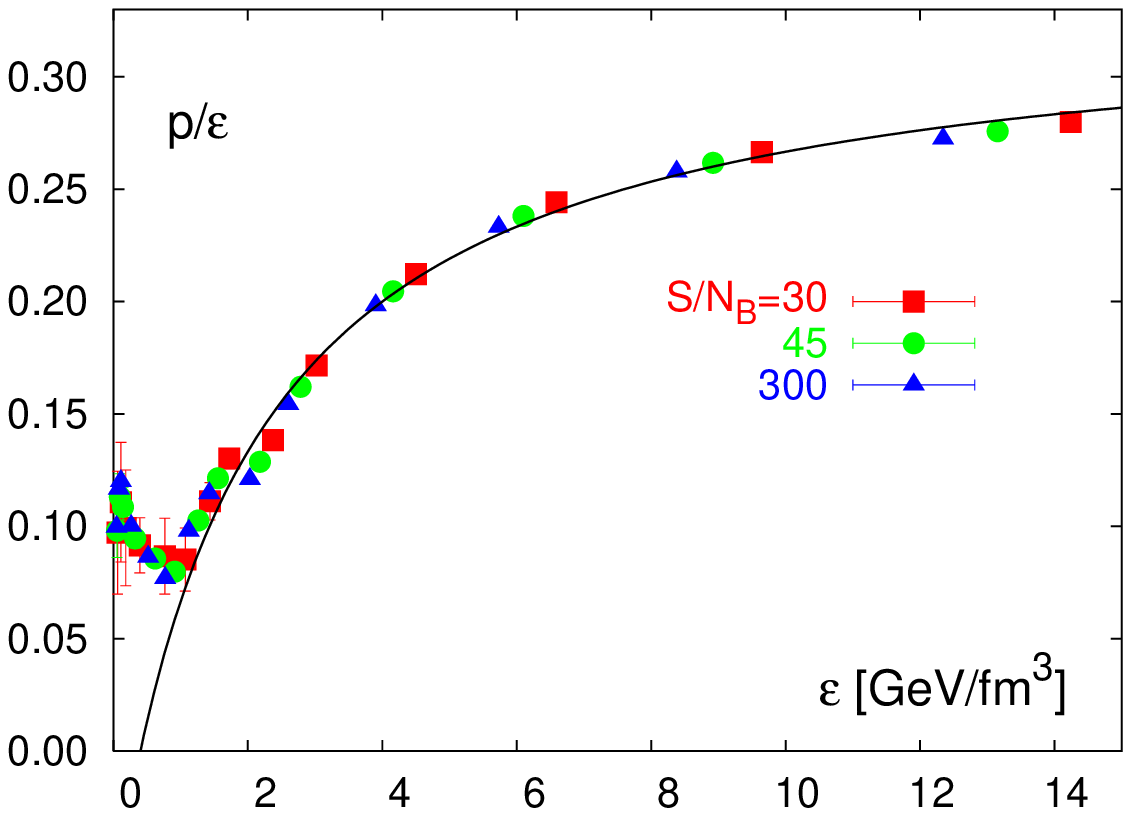, width=7.7cm}
\end{center}
\end{minipage}
\end{center}
\caption{Equation of state on lines of constant entropy per quark number 
versus $T/T_0$ (left) and in physical units using $T_0=175$~MeV to set the 
scale (right). The solid curve in the right hand figure is the 
parametrization of the high temperature part of the equation of state
given in Eq.~\protect{\ref{fit}}.}
\label{fig:soft}
\end{figure}

We now can proceed and calculate energy density and pressure on lines
of constant entropy per baryon number using our Taylor expansion results
up to ${\cal O}(\mu_q^6)$. In Fig.~\ref{fig:eos} we show 
both quantities for the parameters relevant for AGS (FAIR), SPS and RHIC 
energies.
At high temperatures the relevant values of the quark chemical potential
in an ideal quark gas are $\mu_q/T = 0.82$, 0.54 and 0.08, respectively. 
However, as can be 
seen from Fig.~\ref{fig:iso} at temperatures $T\lsim 1.8T_0$ the
required values for the chemical potentials are significantly smaller. 
In particular, for the AGS (FAIR)
energies ($S/N_B=30$) we find $\mu_q/T=0.77$ at $T\simeq 1.8T_0$. 
The $6^{th}$ order 
Taylor expansion thus is still well behaved at these values of the
quark chemical potential. The dependence of $\epsilon$ and $p$ on 
$S/N_B$ cancels to a large extent in the ratio $p/\epsilon$, which
is most relevant for the analysis of the hydrodynamic expansion of dense 
matter. This may be seen by considering the leading ${\cal O}(\mu_q^2)$
correction,
\begin{equation}
\frac{p}{\epsilon} = \frac{1}{3} - \frac{1}{3} 
\frac{\epsilon_0-3p_0}{\epsilon_0} \left(
1 + \left[ \frac{c'_2}{\epsilon_0-3p_0} -
\frac{\epsilon_2}{\epsilon_0} \right] \left( 
\frac{\mu_q}{T} \right)^2 \right) \; .
\label{povere}
\end{equation}
In Fig.~\ref{fig:soft}  we show
$p/\epsilon$ as function of $T/T_0$ (left) as well as function
of the energy density (right). In the latter case we again used 
$T_0 = 175$~MeV to set the scale. The small dependence on $S/N_B$
visible in Fig.~\ref{fig:soft}(left) gets further reduced when
considering the ratio $p/\epsilon$ at fixed energy density 
(Fig.~\ref{fig:soft}(right)). The softest point of the equation
of state is found to be $(p/\epsilon)_{min} \simeq 0.075$ and within
our current numerical accuracy it is independent of $S/N_B$. 
However, the analysis of the temperature regime $T\lsim T_0$ at present
clearly suffers from poor statistics and still needs 
a more detailed analysis.  
We also find that the equation of state for $T_0<T<2T_0$  
is well parametrized by
\begin{equation}
\frac{p}{\epsilon} = \frac{1}{3}\left(1- \frac{1.2}{1+0.5\epsilon\; {\rm fm}^3
/{\rm GeV}}\right) \; .
\label{fit}
\end{equation}
We note, however, that this
phenomenological parametrization is not correct at asymptotically large
temperatures as it is obvious that Eq.~\ref{fit} would lead to corrections
to the ideal gas result $p/\epsilon =1/3$ that are proportional to $T^{-4}$
while perturbative corrections vanish only logarithmically as function of $T$. 

\section{Conclusions}
\label{conclusions}
We have determined Taylor expansion coefficients for the energy and 
entropy densities in 2-flavor QCD at non-zero quark chemical potential.
At present these expansion coefficients have been determined from
calculations performed at one value of the bare quark mass which 
gives a good description of QCD thermodynamics at high temperature.
In this temperature regime lines of constant entropy per
baryon are well described by lines of constant $\mu_q/T$. On these
isentropic lines we have determined the equation of state and find
that the ratio of pressure and energy density shows remarkably little
dependence on the ratio $S/N_B$.

The regime close to and below the transition temperature $T_0$ is not 
yet well controlled in our current analysis. Smaller quark masses and
the introduction of a non-vanishing strange quark mass, kept fixed in
physical units, will be needed to explore this regime as well as the
approach to the 3-flavor high temperature limit in more detail. 

\section*{Acknowledgments}
\label{ackn}
The work of FK and CS has been supported by a contract DE-AC02-98CH1-886 with
the U.S. Department of Energy. FK and EL acknowledge partial support through 
a grant of the BMBF under contract no. 06BI106. SE has been supported by 
the Sumitomo Foundation under grant no. 050408.

\end{document}